**Photoinduced Metal-to-Insulator Transitions in 2D Moiré Devices**

Yiliu Li[1†], Esteban Rojas-Gatjens[1†], Yinjie Guo[2], Birui Yang[2], Dihao Sun[2], Luke Holtzman[3], Juseung Oh[1], Katayun Barmak[3], Cory R. Dean[2], James C. Hone[4], Nathaniel Gabor[5], Eric A. Arsenault[1*], Xiaoyang Zhu[1*]

[1]Department of Chemistry, Columbia University, New York, NY 10027, USA
[2]Department of Physics, Columbia University, New York, NY 10027, USA
[3]Department of Applied Physics and Applied Mathematics, Columbia University, New York, NY 10027, USA
[4]Department of Mechanical Engineering, Columbia University, New York, NY 10027, USA
[5]Department of Physics and Astronomy, University of California, Riverside, CA 92521, USA

**Abstract**

Photoexcitation has been utilized to control quantum matter and to uncover metastable phases far from equilibrium. Among demonstrations to date, the most common is the photoinduced transition from correlated insulators to metallic states; however, the reverse process without initial orders has not been observed. Here, we show ultrafast metal-to-insulator transition in gate-doped $WS_2/WSe_2$ and $WSe_2/WSe_2$ moiré devices using photo-thermionic hole injection from graphite gates. The resulting correlated insulators are metastable, with lifetimes exceeding microseconds. These findings establish an effective mechanism for the ultrafast control of correlated electronic phases in van der Waals (vdW) heterostructures.

† These authors contributed equally to this work.
* Corresponding author. Email: eaa2181@columbia.edu; xyzhu@columbia.edu

Photoexcitation offers a powerful means to probe and control quantum materials. Recent studies have demonstrated photoinduced superconductivity [1–5], ferroelectricity [6–8], and magnetism [9,10]. These transient phases are typically driven by the selective excitation of phonon modes [1,4,6], by strong THz fields [7], or by Floquet engineering [11,12]. Earlier work investigated the ultrafast photoinduced insulator-to-metal transition (IMT) of Mott states following photoexcitation [13–16], resulting in transient Drude-like responses [17]. The reverse processes, photoinduced metal-to-insulator transition (MIT), remains unexplored. The only exceptions are in systems with magnetic order where photoinduced spin-flips destabilize local ferromagnetic order, triggering the transition to a charge/orbital-ordered insulating state [18,19]. Floquet engineering may introduce correlation gaps, but these effects persist during the laser pulse duration of picoseconds or less [20,21]. Most recently, Padma *et al.* reported long-lived metastable states in the cuprate ladder system [22], where symmetry breaking from light-field dressing induces charge transfer from charge reservoir chains to the ladder. Realizing ultrafast switching of a disordered state to an ordered one, with sufficient lifetime, remains elusive.

Here we demonstrate ultrafast MIT using photoinduced charge transfer from graphite electrodes to transition metal dichalcogenide (TMD) moiré superlattices, where an abundance of quantum phases has been observed [23–31]. Previously, we explored photoinduced IMT in $WS_2/WSe_2$ and $MoTe_2/MoTe_2$ bilayer, where photoexcitation across the correlation gap disrupts the correlated insulator phase as detected by exciton sensing [32–34]. In these experiments, carrier doping is achieved electrostatically in a dual-gated structure, where the active TMD layers are encapsulated on top and bottom by hexagonal boron nitride (hBN) dielectric spacers and few-layer graphite (Gr) gates, Fig. 1a. While photoexcitation targets the correlated states in the moiré TMD bilayers [32–34], the graphite gates are unavoidably photoexcited. Excitation of the Gr-gate electrodes has been shown to generate phonon wavepackets (strain waves) that propagate from the graphite gates through the hBN spacer to arrive at active TMD layers with characteristic delay times [35,36]. At sufficiently high excitation densities, excitation in the Gr-gate electrodes results in photo-thermionic emission [37,38], which we show here as a source for ultrafast carrier injection into the TMD layers (Fig. 1b) to result in metastable state persisting up to tens of microseconds [39,40].

We establish photoinduced MIT in $WS_2/WSe_2$ and $WSe_2/WSe_2$, illustrated in Fig. 1a, where steady-state electron or hole density is controlled through electrostatic gating. We focus on the $WS_2/WSe_2$ heterobilayer (twist angle $\theta = 60.0 \pm 0.5^o$) in the main text and present results for the $WSe_2/WSe_2$ homobilayer ($\theta = 58.1 \pm 0.1^o$), as well as detailed characterization of the photo-thermionic emission mechanism, in a companion paper [39]. See Supporting Information Figs. S1 and S2 for optical images of the devices and selected characterizations [40]. We chose the $WS_2/WSe_2$ heterostructure because the integer filling insulators feature high critical temperatures ($T_c$ ~150 K) [23,25,41]. At sufficient excitation density, the high-energy tail of hot holes in Gr-electrodes reaches the valence band edge of the hBN spacer. The resulting photo-thermionic emission [37,38] serves to transiently dope the TMD moiré layers, Fig. 1b, as detailed separately [39].

We apply time-resolved reflectance to probe the melting and formation of correlated states. The pump photon energy of 1.46 eV selectively excites the correlated states, but is below the optical gap of the TMDs. We use the $WSe_2$ moiré exciton as a sensor for the quantum phases of doped charge [32,33,42]. Sensing the $WS_2$ exciton and trion spectral regions shows similar results, but with a lower signal level (Fig. S3-S5), likely a result of photonic effect in this wavelength range. The exciton oscillator strength ($f$) scales inversely with the Bohr radius and, thus, the effective dielectric constant ($\varepsilon_{eff}$), $f \propto {\varepsilon_{eff}}^{-\alpha}$, where the exponent $\alpha (\geq 1)$ depends on the detailed form of the Coulomb potential [43,44]. The opening of a correlation gap or pseudo gap results in a decrease in $\varepsilon_{eff}$ and the corresponding increase in $f$, giving rise to distinct signatures for photo-induced IMT (Fig. 1c) and MIT (Fig. 1d).

A reflectance spectrum features the characteristic derivative shape as photon energy increases across a resonance, shown in Fig. 1e for an insulator (red) and a metallic (blue) state; with the former possessing higher oscillator strength. In transient reflectance $\Delta R/R_0$, where $\Delta R = R(\Delta t) - R_0$; $R(\Delta t)$ and $R_0$ are the reflectance spectrum at $\Delta t$ and without pump pulse (static), respectively. For IMT, the pump-induced reduction in $f$, i.e., photobleach (PB), gives a $\Delta R/R_0$ spectrum with a characteristic flip in sign, Fig. 1f, as compared to the static spectrum. In photoinduced MIT, the increase in oscillator strength $f$, commonly referred to as photon-induced absorption (PIA), yields a $\Delta R/R_0$ spectrum of derivative shape with the same sign as that of the static reflectance spectrum, Fig. 1g. In the following, we take the opposite signs in $\Delta R/R_0$ spectra as signatures for IMT and MIT. Note that previous experiments on exciton sensing of correlated states in TMD moiré bilayers have shown that photo-induced IMT or doping-induced MIT results in mainly changes in exciton oscillator strength, without major changes in resonance energy [32–34]. This is likely due to the well-known cancellation effort between bandgap renormalization and dielectric screening of exciton binding energy in two dimensional semiconductors [45].

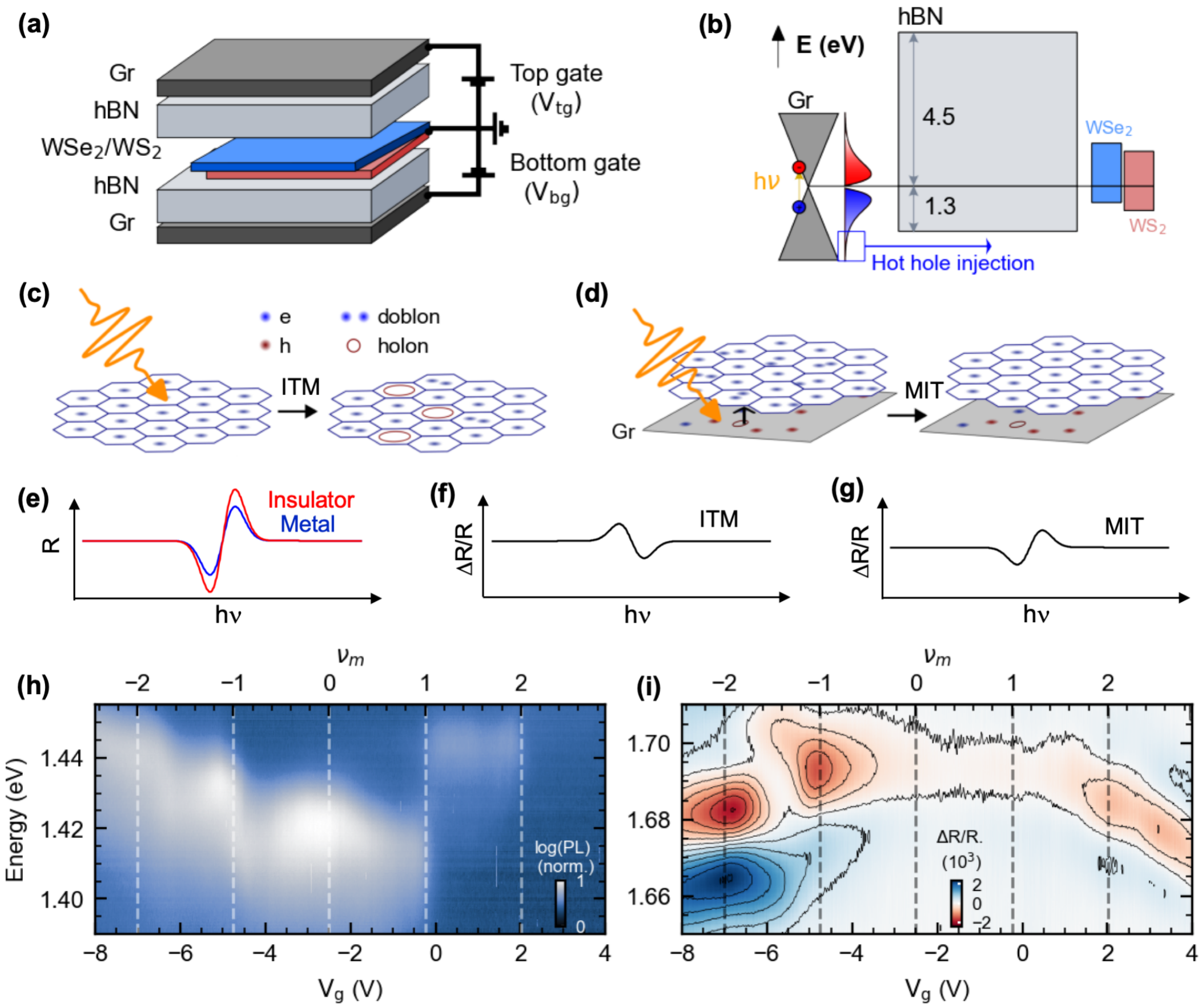

**Figure 1.** (a) Schematic of the device showing the top ($V_{tg}$) and bottom ($V_{bg}$) gates consisting of few-layer graphite (Gr) and hexagonal boron nitride (hBN) encapsulating the $WSe_2/WS_2$ heterobilayer with a twist angle close to 60°. (b) Schematic of band alignment and thermionic hole injection mechanism. (c) Schematic of photoinduced insulator-to-metal transition (IMT). Excitation across the correlated gap induces disordering, thus increasing the dielectric screening and resulting in a photobleach-like response. (d) Schematic of photoinduced metal-to-insulator transition (MIT). Pump pulse excites hot electrons/holes in the few-layer graphite gate (gray layer), and the resulting hot hole population is injected into the $WS_2/WSe_2$ heterobilayer, thus transiently modifying electron density. The expected response in transient reflection experiments is summarized in schematic diagrams: (e) a comparison of reflectance spectra (R) from a metal (blue) and an insulator (red); transient reflectance spectra (ΔR/R) for (f) IMT and (g) MIT. (h) Steady-state photoluminescence spectrum as a function of gate voltage ($V_g = V_{tg} = V_{bg}$). We used a continuous-wave laser with an energy of 2.33 eV (42 nW/µm²) to excite the sample. (i) Transient reflectance as a function of gate voltage measured at pump-probe delay Δt = 10 ps. Pump photon energy, hν = 1.46 eV, rep-rate 400 kHz, pulse energy density 21 µJ/cm$^2$. The broadband probe hν = 1.55-1.77 eV (rep-rate 400 kHz, pulse energy density 15 µJ/cm$^2$) was set to cover the $WSe_2$ moiré exciton. Unless otherwise specified, the sample temperature was kept at 9 K.

To characterize the sample, we monitor the photoluminescence of the $WS_2/WSe_2$ interlayer exciton as a function of gate voltage ($V_g$), which sensitively detects increases in exciton oscillator strength upon formation of insulator states. Fig. 1h shows PL spectra as a function of $V_g$, from which we observe four distinct states with lower $\varepsilon_{eff}$, corresponding to electron ($\nu_m$ = 1, 2) and hole ($\nu_m$ = -1, -2) correlated insulators. Details on calibrating the moiré lattice filling factor, $\nu_m$, can be found in Methods. Due to a built-in electric potential from intrinsic doping of the component materials, the charge neutrality point $\nu_m$ = 0 is not at $V_g$ = 0. Fig. 1i shows doping-dependent transient reflectance at a low pulse energy density ρ = 21 μJ /cm$^2$ and a selected pump-probe delay Δt = 10 ps. We observe clear PB features throughout the doping region, particularly for integer filling factors, $\nu_m$ = -1, -2, 2. The melting and recovery of the correlated insulators are observed in the entire experimental time window, Fig. S6-10.

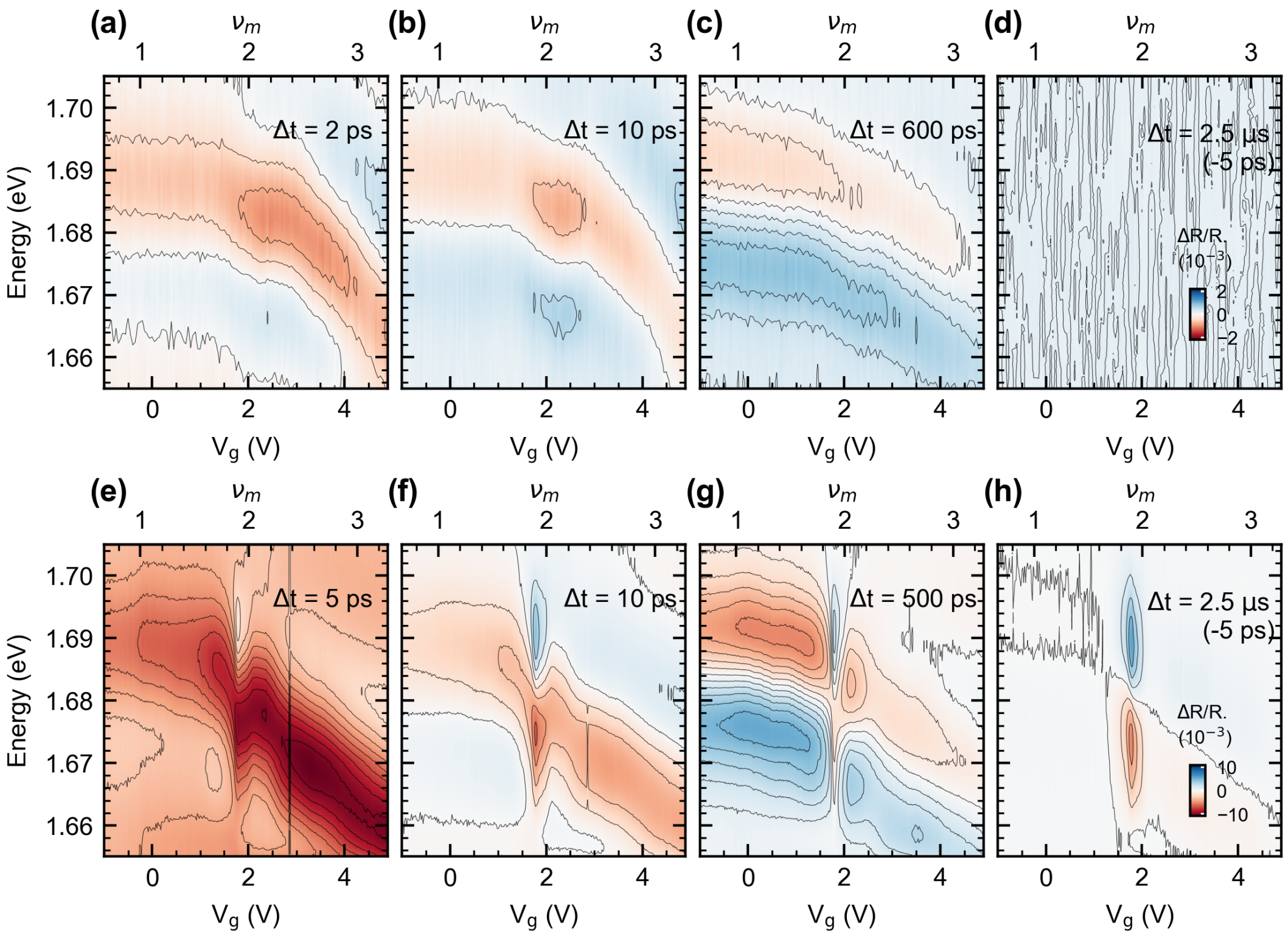

**Figure 2.** Transient reflectance as a function of gate voltage at pump-probe delays: Δt = 5, 10, 600, 2.5 μs (-5 ps), with a fluence of 13 μJ/cm$^2$, and the probe spectrum was set to cover the $WSe_2$ moiré exciton, corresponding to panels (a), (b), (c), and (d). Transient reflectance as a function of gate voltage at pump-probe delays: Δt = 5, 10, 600, 2.5 μs (-5 ps), with a fluence of 127 μJ/cm$^2$, and the probe spectrum was set to cover the $WSe_2$ moiré exciton, corresponding to panels (e), (f), (g), and (h).

Now we demonstrate the distinct changes in photo-induced effects as excitation power density increases. Figs. 2a-d show reflectance spectra on the electron doping side at Δt = 2 ps, 10 ps, 600 ps, and 2.5 μs, respectively, at the low pulse energy density $\rho$ = 13 μJ /cm$^2$. The PB feature is observed in the entire doping range, indicating the disruption of correlation, including the photo-induced IMT for the $\nu_m$ = 2 insulator state. The PB feature disappears at Δt >> 600 ps as the excitation dissipates and correlation recovers. Note that Δt = 2.5 μs, equivalent to the delay from the prior pulse determined by the laser repetition rate (400 kHz), corresponds to a negative delay (Δt = -5 ps) in the experiment.

In stark contrast to the results at low ρ in Figs. 2a-d, we observe a reverse of sign in the transient reflectance spectra around $V_g$ = 2 V at ρ = 127 μJ /cm$^2$, Figs. 2e-h. The PIA feature is visible at Δt as short as 5 ps, Fig. 2e, sitting on a background of the ultrafast pump-probe response of the graphite gate and the PB feature from melting/recovery of correlation. The signature of photo-induced MIT as PIA becomes clearly resolved on the melting/recovery background of PB signal at Δt = 10 and 600 ps and becomes the only feature at Δt = 2.5 μs. This PIA feature becomes more pronounced with increasing ρ (see Fig. S11 and reference [39]). Spectrally resolved dynamics at different doping levels can be found in the Supporting Information, Figs. S12-S15.

Following excitation and ultrafast electronic thermalization in the graphite gate, a hot electron distribution is reached on sub-picosecond time scales [46,47]. The electronic temperature ($T_e$) scales with photoexcitation density [48], $T_e \propto \rho^{\beta}$, where the exponent $\beta$= 1/3 at the 2D limit. This leads to a nearly exponential increase in the high-energy tail in the hot electron distribution as $\rho$ increases, with sufficient energy to reach the valence band maximum (VBM) of the hBN barrier and resonantly transports to the TMD bilayer. The energy offset between the Fermi level of graphene and the VBM of hBN is approximately 1.3 eV, much smaller than the 4.5 eV offset to the conduction band minimum (CBM) [38,49]. This energetic offset enables the injection of holes, but not electrons. Moreover, a positive $V_g$ on the electron doping side favors hole injection while the negative $V_g$ on the hole doping side inhibits it. In the companion work, we establish the photoinduced hole injection mechanism [39]. The photoinjected holes recombine with the electrostatically doped electrons in the $WS_2$/$WSe_2$ heterobilayer, thus transiently reducing the electron density from that of a liquid state at $\nu_m$ > 1 to a correlated insulator, likely the robust $\nu_m$

= 1 Mott state. Supporting this interpretation, we find that the sharp PIA feature appears at higher static $\nu_m$ values with increasing excitation fluence; detailed establishment of hole injection mechanism can be found in our companion work [39].

While the correlated insulator is formed on ultrafast timescales, the subsequent discharge of is much slower and limited by the large contact resistances and corresponding RC constants (R: resistance; C: capacitance) [50]. The persistent charge imbalance lasts longer than the 2.5 μs inter-pulse time spacing, leading to PIA signal at pre-time zero, Fig. 2h. The slow discharge provides metastability to the photoinduced insulator states. This metastability allows us to exclude an alternative interpretation – the observed PIA originating from a mixed electron/exciton insulator [51–53]. This scenario is unlikely because the interlayer exciton lifetimes of nanoseconds [45,54,55] are three orders of magnitude lower than the microsecond lifetime observed here.

We now turn to the initial step of hole-injection and photoinduced MIT. As detailed in Supporting Information, Figs. S16, the initial time-dependent ΔR/R response is independent of doping level in the TMD moiré structure and can be attributed to ultrafast hot electron dynamics in the Gr electrodes [46,47], with an initial single-exponential lifetime of 220 fs. We can obtain the dynamic responses of correlated states in the moiré TMD bilayer at $\nu_m \neq 0$, Fig. 3, by subtracting that at $\nu_m = 0$. At the $\rho = 13$ μJ/cm$^2$, Fig. 3a, all correlated states on either electron or hole doping sides feature the characteristic electronic melting and recovery dynamics [32–34], in addition to high frequency modulation from the delayed arrival (≥ 10 ps) of acoustic phonon wavepackets launched at the Gr electrodes [35]. The solid curves show fits to the melting and recovery dynamics, with extracted lifetimes summarized in Supporting Information Table 1. At the high excitation fluence of $\rho = 127$ μJ/cm$^2$, Fig. 3b, the time profile of the PIA feature attributed to photoinduced MIT is in stark contrast to those featuring an initial decrease, i.e., IMT, followed by recovery at $\nu_m$ = -1 and -2. Instead, the PIA feature shows an initial rise to nearly a plateau. Single exponential fit (solid curve on top of the data) gives an initial rise time of $\tau_i = 1.1 \pm 0.2$ ps, attributed to hole injection from the Gr electrodes and the corresponding photoinduced MIT. This value is consistent with previous reports on hole injection in Gr/hBN/Gr heterostructures [38,56]. Due to a time resolution of approximately 200 – 300 fs, processes occurring on faster timescales cannot be resolved in the present measurements. We point out that the rise time of the photo-induced MIT, $\tau_i$ = 1.1±0.2 ps, is one order of magnitude shorter than the arrival time (~ 10 ps) at the moiré interface of the acoustic phonon wavepacket (strain wave). The latter can induce re-melting of correlated phases on much longer time scale (~100 ps) due to the reheating of the electronic degrees of freedom [34].

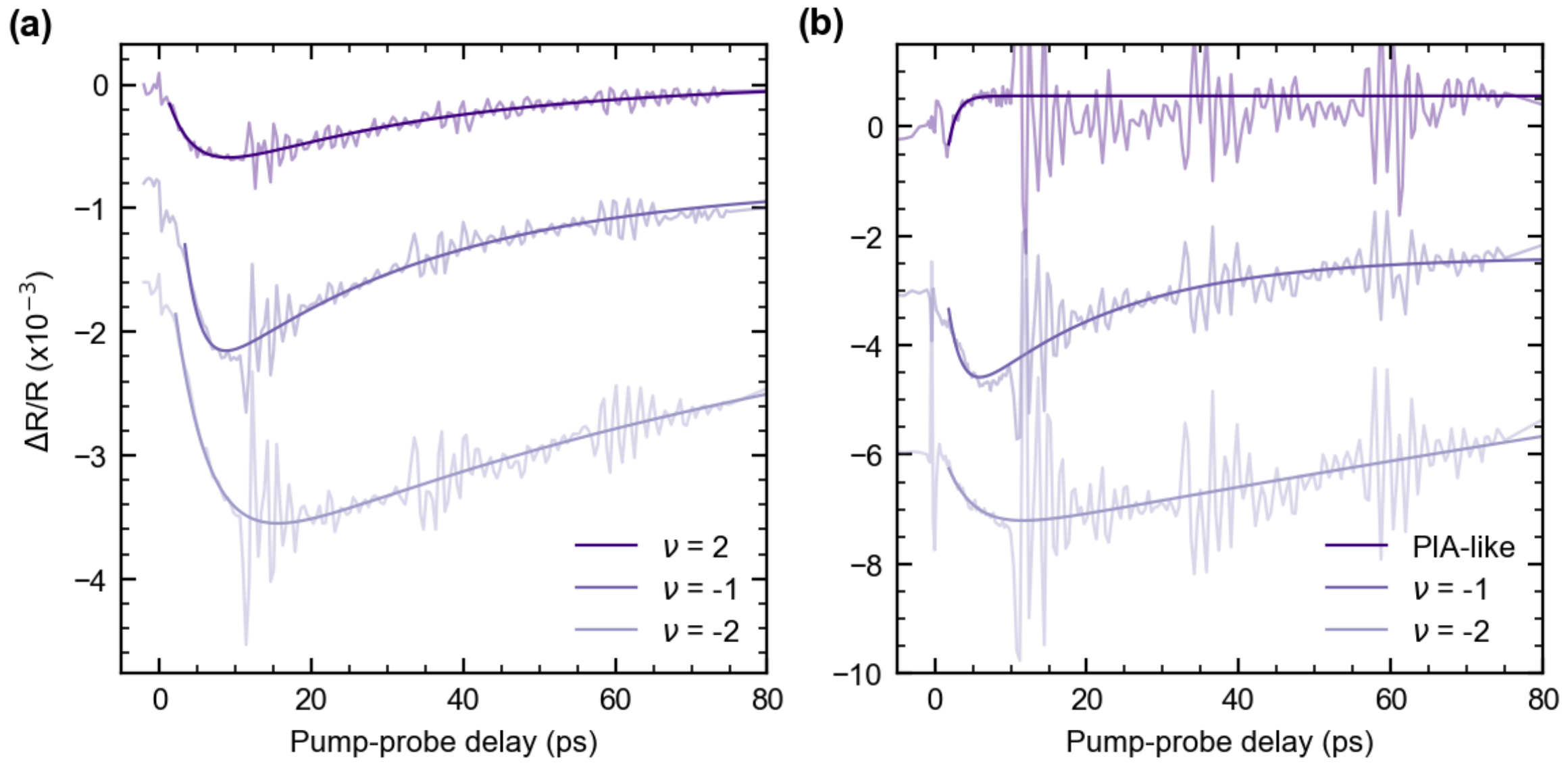

**Figure 3.** Dynamic responses of correlated states at low and high pump fluence, (a) and (b) respectively. Dynamics of correlated states $\nu_m$ =2, -1, -2 with pump energy of 1.46eV and fluence of 13 μJ/cm$^2$, by subtracting the corresponding response at charge neutral state. Dynamic responses of correlated states $\nu_m$ = -1, -2, and states showing PIA feature with pump energy of 1.460eV and fluence of 127μJ/cm$^2$, by subtracting the corresponding response at charge neutral state. The time trace was obtained by averaging a 10 meV energy window centered at the maximum of the high-energy feature.

Supporting photoinduced MIT, we show the presence of a critical temperature characteristic of a correlated insulator state formed transiently from hole injection. Figure 4 shows the transient reflectance spectra at $\rho$ = 318 μJ /cm$^2$ and Δt = 2.5 μs when only the metastable state remains. The PIA feature consists of two distinct components: i) a sharp peak attributed to the photoinduced MIT, and ii) a broader feature at higher gate voltages. The MIT peak diminishes with increasing temperature and becomes negligible at the highest temperature, 163 K, probed here. This temperature dependence is consistent with the $T_c$ ~ 150 K for the Mott insulator states in electron or hole-doped $WS_2/WSe_2$ moiré structure. In contrast, the broad background feature persists at temperatures above the critical temperature of any correlated phases in this system. In the case of the $WSe_2/WSe_2$ homobilayer [39], we measured this feature at a temperature as high as 295 K, as shown in Fig. S17. Therefore, the broad background cannot be related to a correlated phase, but instead reports on an increase in the oscillator strength. At higher gate voltages, the increased contrast between the densities of electrostatically doped electrons and photoinjected holes leads to reduced screening, as electron–hole recombination transiently decreases the total carrier density.

Concluding remarks. We present an ultrafast pathway for charge injection into moiré materials in vdW devices and demonstrate a rare example of photoinduced MIT in gate-doped $WS_2/WSe_2$ and $WSe_2/WSe_2$ moiré devices. The photo-thermionic hole injection occurs on picosecond time scales, thus allowing the ultrafast switching of quantum phases not possible with electrostatic gating. The resulting transient states are metastable on microsecond time scales. While we focus on the photoinduced MIT here, such an ultrafast charge injection mechanism may be explored for the tuning and interrogation of a variety of moiré quantum phases, and for applications in fast electronics, such as a flash memory [57]. Furthermore, integrating ultrafast optics with advanced device architectures may allow the exploration for light-driven, programmable quantum phases in vdW systems.

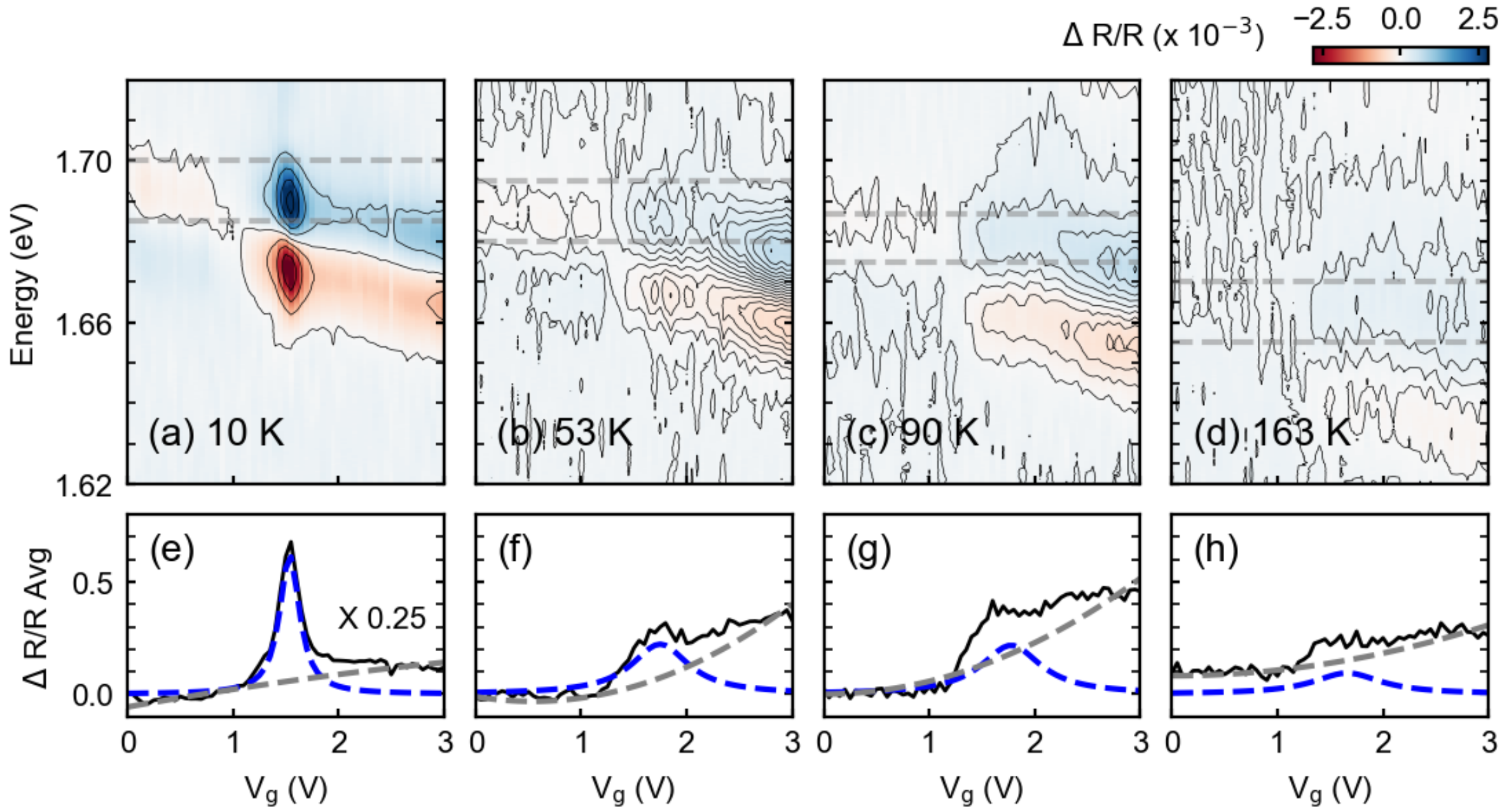

**Figure 4.** Temperature-dependent (10 K, 53 K, 90 K, 165 K, as labeled for each column) transient reflectance as a function of gate voltage at a pump-probe delay of $\Delta t$ = 2.5 μs (-5 ps), and the corresponding averaged integrated reflectivity. The blue line corresponds to Lorentzian component while the grey line corresponds to the background. The sample is only gated with bottom gate during the temperature-dependent measurements.

## Acknowledgments

This work was supported by the US Department of Energy Office of Basic Energy Sciences (DOE-BES) under award DE-SC0024343. Device fabrication and characterization were supported by the National Science Foundation grant DMR-2505586. Development of the photo-thermionic emission mechanism was supported by Department of Defense Multidisciplinary University Research Initiative (MURI) grant number W911NF2410292. E.R-G. acknowledges support from the Columbia Quantum Initiative Postdoctoral Fellowship. E.A.A. gratefully acknowledges support from the Simons Foundation as a Junior Fellow in the Simons Society of Fellows (965526).

**Author contributions**

Y.L. prepared the twisted bilayer devices. Y.L., E.R-G., and E.A.A. carried out the optical measurements and analysis. L.H. prepared the TMD single crystals under the supervision of J.H. and K.B. Y.G., B.Y., D.S and J.O. assisted with the preparation of the devices under the supervision of CD. N.G. and X.-Y.Z formulated the photo-thermionic emission mechanisms. X.Y.Z. supervised the project. This manuscript was prepared by Y.L, E.R-G., E.A.A., and X.Y.Z. in consultation with all other authors. All authors read and commented on the manuscript.

**SM: Photoinduced Metal-to-Insulator Transitions in 2D Moiré Devices**

Yiliu Li, Esteban Rojas-Gatjens, Yinjie Guo, Birui Yang, Dihao Sun, Luke Holtzman, Juseung Oh, Katayun Barmak, Cory R. Dean, James C. Hone, Nathaniel Gabor, Eric A. Arsenault, Xiaoyang Zhu

## Methods

### Device fabrication

$WSe_2$ monolayers were mechanically exfoliated from flux-grown $WSe_2$ bulk crystals [1]. $WS_2$ monolayer was mechanically exfoliated from commercially available bulk crystals (2D Semiconductors). To fabricate the $WS_2/WSe_2$ heterostructure, we first determined crystal orientations of $WSe_2$ and $WS_2$ monolayers using polarization-resolved second harmonic generation (SHG). We stacked the monolayers using the dry-transfer technique with a polycarbonate stamp. To distinguish between R-stack and H-stack samples, the SHG measurement was carried out in the heterolayer region after the device was fabricated, with the results compared to those of individual monolayers, Figure S1. The tWSe2 homobilayer was stacked by the dry transfer "tear and stack" technique with a polycarbonate stamp. The moiré period was confirmed by piezo force microscopy (PFM, Figure S2). The twist-angle uncertainty is extracted from the fitting error of the polarization-resolved SHG measurements for the $WS_2/WSe_2$ heterostructure, and from the filling-factor calibration uncertainty in static reflectance gate scan for the $tWSe_2$ bilayer. The sample was grounded via a few layer graphite (Gr) contacts connected to the heterolayer and single-crystal h-BN dielectrics, and graphite gates are used to encapsulate the device and provide control of the carrier density and displacement field (if desired) via external source meters. All electrodes are defined with electron beam lithography and made of a three-layer metal film of Cr/Pd/Au (3 nm/17 nm/60 nm).

### Steady-state photoluminescence spectroscopy

The steady state photoluminescence (PL) spectra were obtained with a continuous wave 532nm laser excitation. The excitation power was set at 42 $nW/\mu m^2$ on the sample. The PL signal was spectrally filtered from the laser through a long-pass filter before dispersal with a spectrometer and detected on a CCD camera.

Time-resolved differential reflectance spectroscopy

We carried out ultrafast transient reflectance measurements using a femtosecond laser system (Pharos, Light Conversion) operating at 1030 nm with an 89 fs pulse duration, 400 kHz repetition rate, 10 W average power, and 20 μJ pulse energy. The laser output was split into pump and probe beams. To generate the probe, a portion of the fundamental beam was focused into a YAG (Yttrium Aluminum Garnet) crystal to produce a white-light continuum. This continuum was spectrally filtered by a 700 nm long-pass and a 800 nm short-pass filter to isolate the energy window corresponding to the lowest $WSe_2$ moiré exciton. The pump beam was sent through a tunable optical parametric amplifier (Orpheus-NEO, Light Conversion, 315–2700 nm) to produce the desired excitation wavelength. It then passed through a motorized delay stage to control the pump-probe time delay (Δt) and an optical chopper for modulation between pump-on and pump-off conditions. Pump and probe beams were recombined and focused onto the sample through a 100×, 0.75 NA objective lens. The diameter of pump and probe spot sizes at the sample were ~1.38 μm and ~1.22 μm, respectively. The reflected probe light was collected by the same objective, filtered to remove residual pump light, dispersed by a spectrometer, and detected using a CCD array (Blaze, Princeton Instruments). The transient reflectance signal (ΔR/R) was calculated from the difference between the pump-on and pump-off spectra at various delay times.

Assignment of the filling factors

The charge density, in the $WS_2/WSe_2$ heterostructure, controlled by the applied gate voltages ($V_t$ and $V_b$ for the top and bottom gates, respectively), was determined based on the parallel-plate capacitor model: $n = \frac{\varepsilon\varepsilon_0 \Delta V_t}{d_t} + \frac{\varepsilon\varepsilon_0 \Delta V_b}{d_b}$, where $\varepsilon \approx 3$ is the out-of-plane dielectric constant of hBN, $\varepsilon_0$ is the permittivity of free space, $\Delta V_i$ is the applied gate voltage, and $d_i$ is the thickness of the hBN spacer. The moiré density, $n_0$, is determined from the moiré lattice constant, $a_M$ according to $n_0 = \frac{2}{\sqrt{3}a_M^2}$. The filling factor, $\nu = \frac{n}{n_0}$, was calibrated against gate-dependent steady state and transient reflectance measurements. The thickness of the top and bottom hBN spacers was $d_t \approx 41.3$ and $d_b \approx 40.3$ nm, respectively.

**Supplementary notes**

Transient response from the $WS_2$ exciton: In Figure S3, we measured the transient response over the wavelength range covering both the $WS_2$ and the $WSe_2$ excitonic states. It can be clearly observed that the $WSe_2$ exciton has a stronger transient response than the $WS_2$ exciton state. We attribute this to a mixture of a photonic effect caused by the substrate, which favors the reflection of the light in the $WSe_2$ exciton range, and a lower response of $WS_2$ to the effective dielectric constant due to a more tightly bound exciton state. Interestingly, we remeasured the samples focusing on the $WS_2$ response, see Figure S4, and observed a transient photoinduced response observed at low excitation density and without any signs of metastability. This feature can be attributed to an enhancement of the trion oscillator strength as the free carrier become available after melting. From the dynamic response, Figure S5, it can be seen the trion oscillator strength follows the melting and recovery of the correlated states sensed by the $WS_2$ exciton.

Data analysis of the dynamics of the $WSe_2$/$WS_2$ moiré device: To analyze the transient reflectance spectra and isolate the melting, recovery, and formation dynamics, we averaged the spectra at the maximum for each feature in a ~0.01 eV window, see the raw data in Figures S6-S15. In addition to the coherent spike, the initial time-dependent ΔR/R response is dominated by ultrafast hot electron dynamics in the Gr gates and independent of doping level in the TMD moiré structure. The dynamic profile at $\nu_m = 0$ is used as representing the response from Gr gates; this background is subtracted from all other dynamic profiles at $\nu_m \neq 0$. We analyze the background subtracted data profiles at $\nu_m \neq 0$ by fitting to a multiexponential of the form of $(\sum A_i e^{-t/t_1} + C)$. The fits are summarized in Fig. 3 in the text, with fitting parameters summarized in Table S1.

**Table S1. Fitting parameters of background-subtracted kinetic profiles (Fig. 3a and 3b).**

| Sample condition | $A_1$ | $t_1$ (ps) | $A_2$ | $t_2$ (ps) | C |
|---|---|---|---|---|---|
| low fluence Figure 3a $(\nu = -2)$ | 3.5 ± 0.2 | 4.5 ± 0.4 | -2.8 ± 0.1 | 96 ± 5 | 0.31 ± 0.03 |
| $(\nu = -1)$ | 6 ± 1 | 2.1 ± 0.3 | -1.9 ± 0.1 | 31 ± 3 | 0.01 ± 0.06 |
| $(\nu = 2)$ | 1.0 ± 0.1 | 3.5 ± 0.5 | -0.91 ± 0.05 | 32 ± 4 | 0.02 ± 0.03 |
| high fluence Figure 3b $(\nu = -2)$ | 2.1 ± 0.9 | 3.7 ± 1 | -11* | 435* | 10* |
| $(\nu = -1)$ | 5 ± 1 | 1.9 ± 0.7 | -3.3 ± 0.3 | 19 ± 3 | 0.6 ± 0.1 |
| PIA | - | - | -4.4 ± 0.1 | 1.1 ± 0.2 | 0.6 ± 0.1 |

*At longer delay times (< 100 ps), there is a second process occurring, phonon heating); the long-time dynamics is not properly included here and can be found in ref. [32] [33] [35].

## Supplementary figures

Device fabrication and characterization

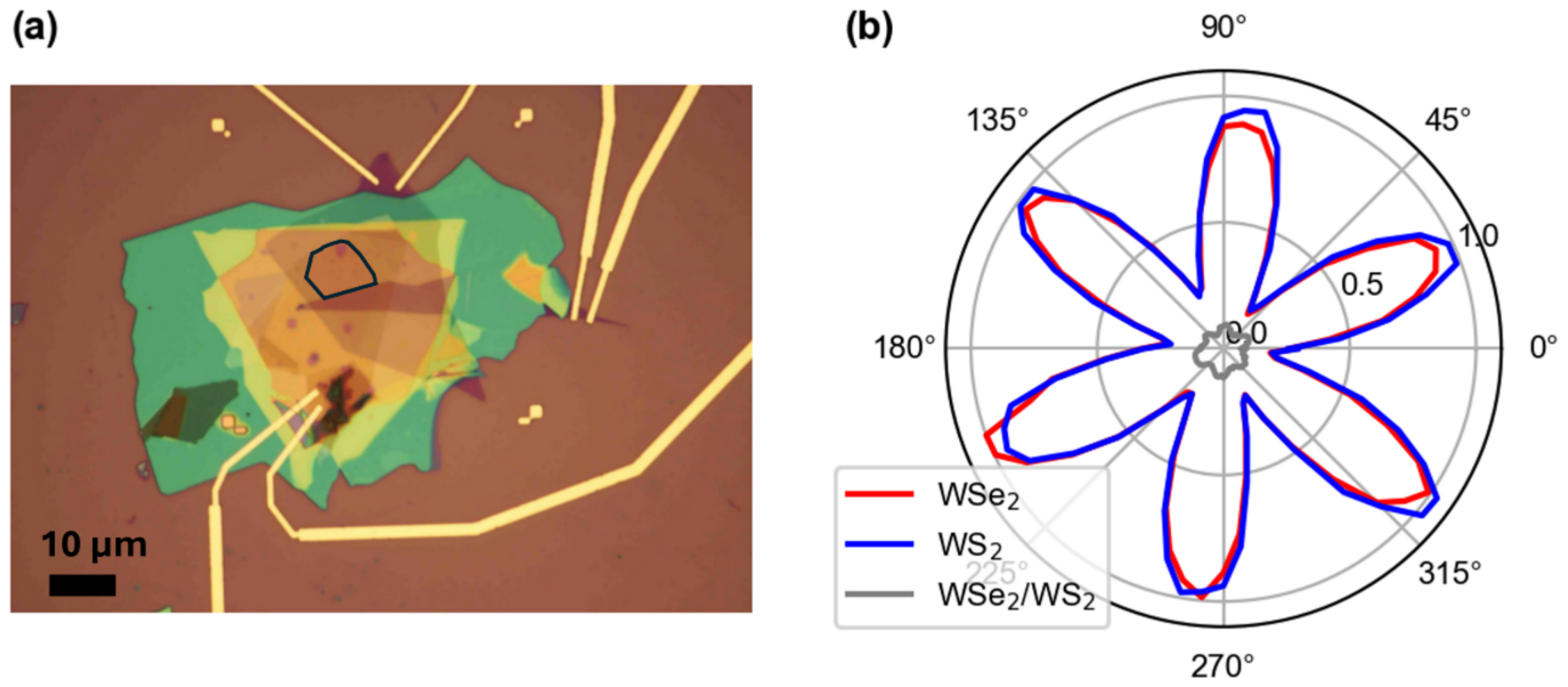

**Figure S1.** (a) Optical image of the $WSe_2/WS_2$ moiré device measured. The enclosed area corresponds to the active bilayer region. (b) Polarized second harmonic spectroscopy for the individual monolayers ($WSe_2$: red, $WS_2$: blue) and the overlapped region ($WS_2/WSe_2$: gray).

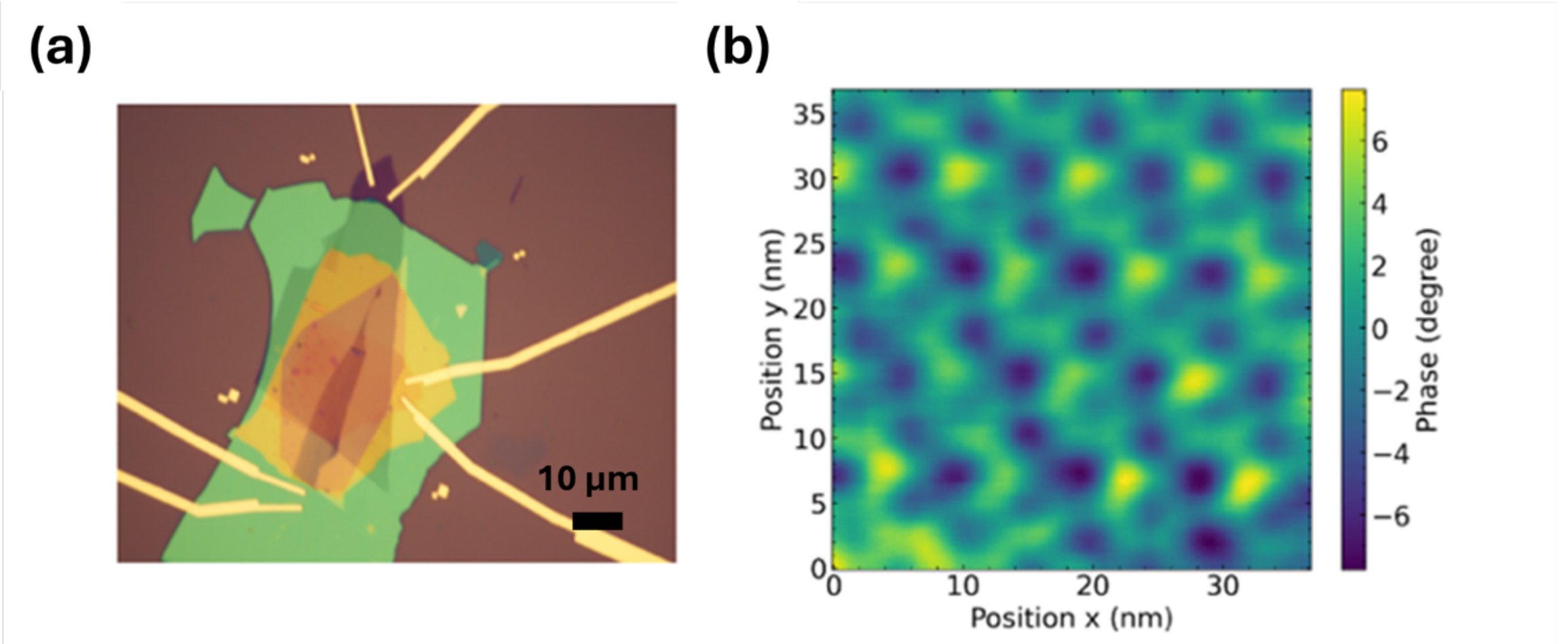

**Figure S2.** (a) Optical image of the $tWSe_2$ moiré device measured. (b) Piezo force microscope image of the moiré superlattice .

<u>Transient response probed by the $WS_2$ exciton</u>

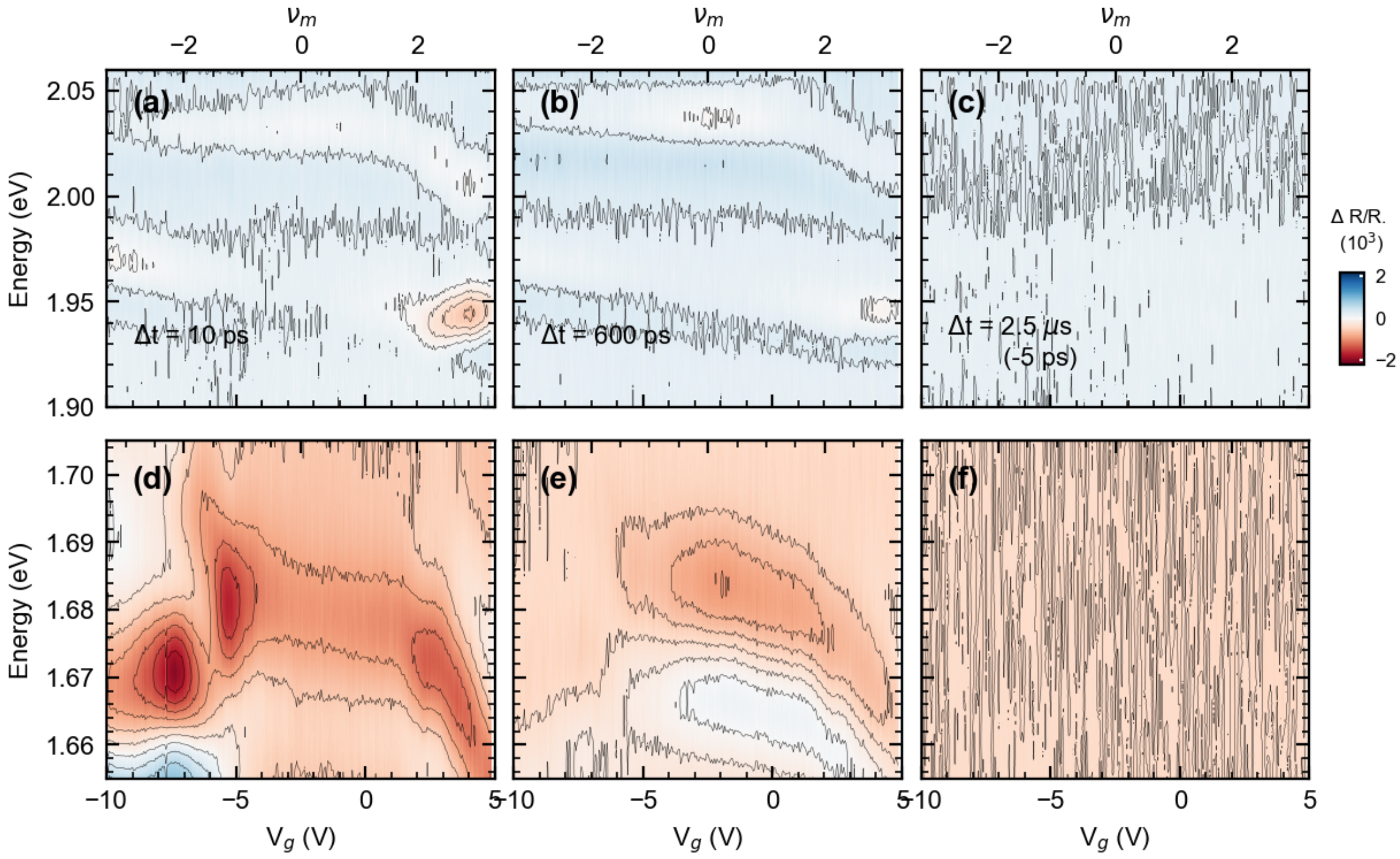

**Figure S3.** Transient reflectance as a function of gate voltage at pump-probe delays: Δt = 10, 600, and 2.5 μs (-5 ps). We used a pump excitation energy of 1.46 eV with a fluence of 42 $\mu J/\mu m^2$, and the probe spectrum was set to cover both the $WS_2$ (a-c) and $WSe_2$ (d-f) exciton.

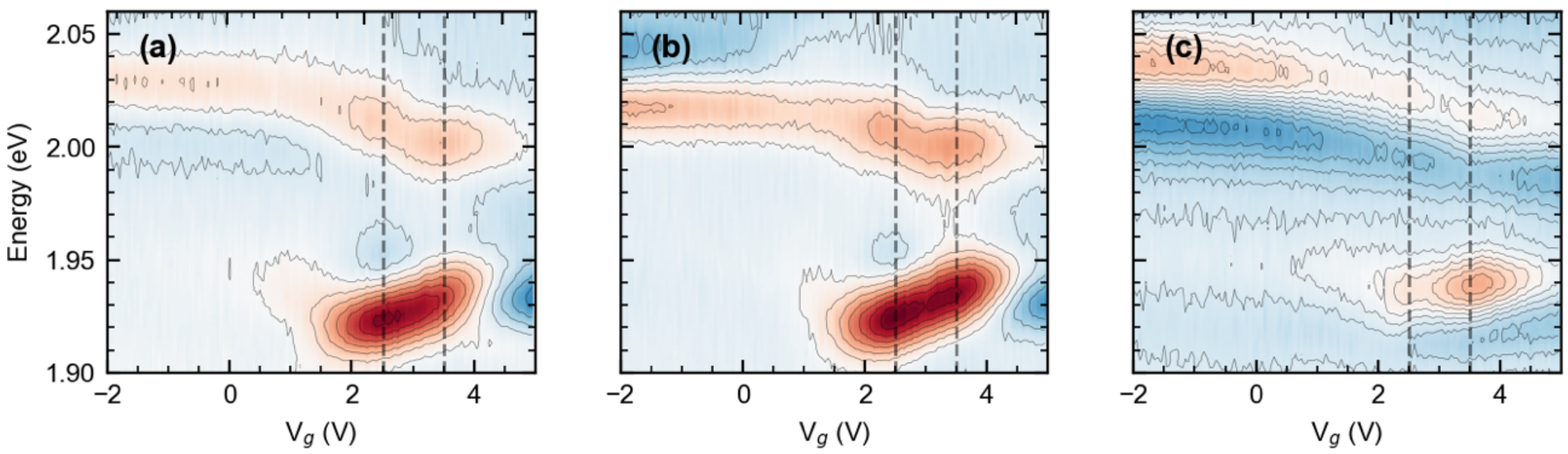

**Figure S4.** Transient reflectance as a function of gate voltage using $WS_2$ as sensor at pump-probe delays: Δt = 2, 10, 400 ps. We used a pump excitation energy of 1.46 eV with a fluence of 42 $\mu J/\mu m^2$, and the probe spectrum was set to cover both the $WS_2$ exciton and trion resonances.

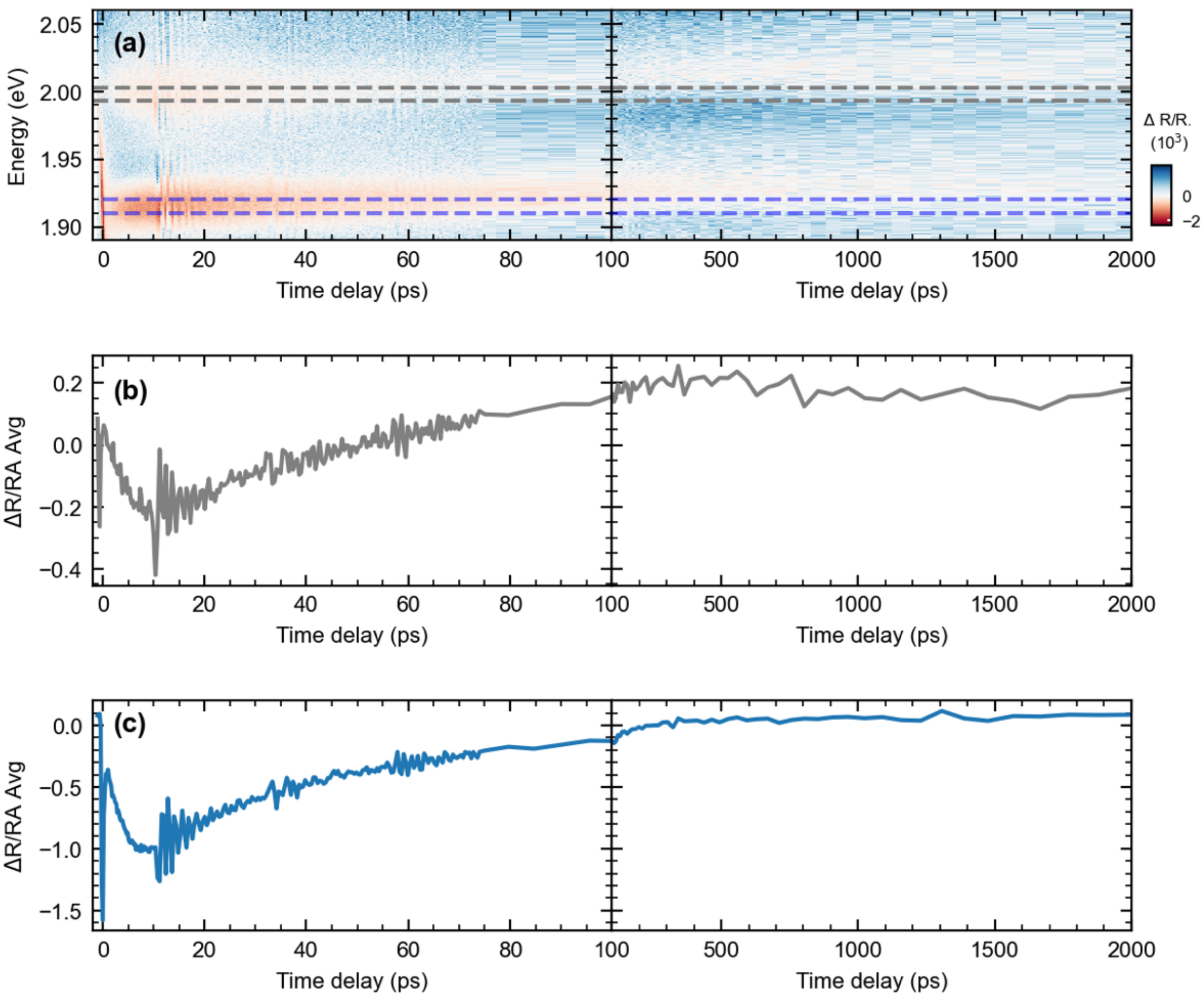

**Figure S5.** (a) Transient reflectance spectra at a fixed gate voltage (2.5 V) as a function of pump-probe delays measured while probing the $WS_2$ exciton and trion resonance. We show cuts at specific energies (integrated over a 10 meV range), corresponding to (b) 1.915eV and (C) 1.998 eV. These traces where measured with the same conditions as the gate scan shown in Figure S4.

Raw dynamics for the $WSe_2/WS_2$ heterostructure at low fluence 21 uJ/cm$^2$

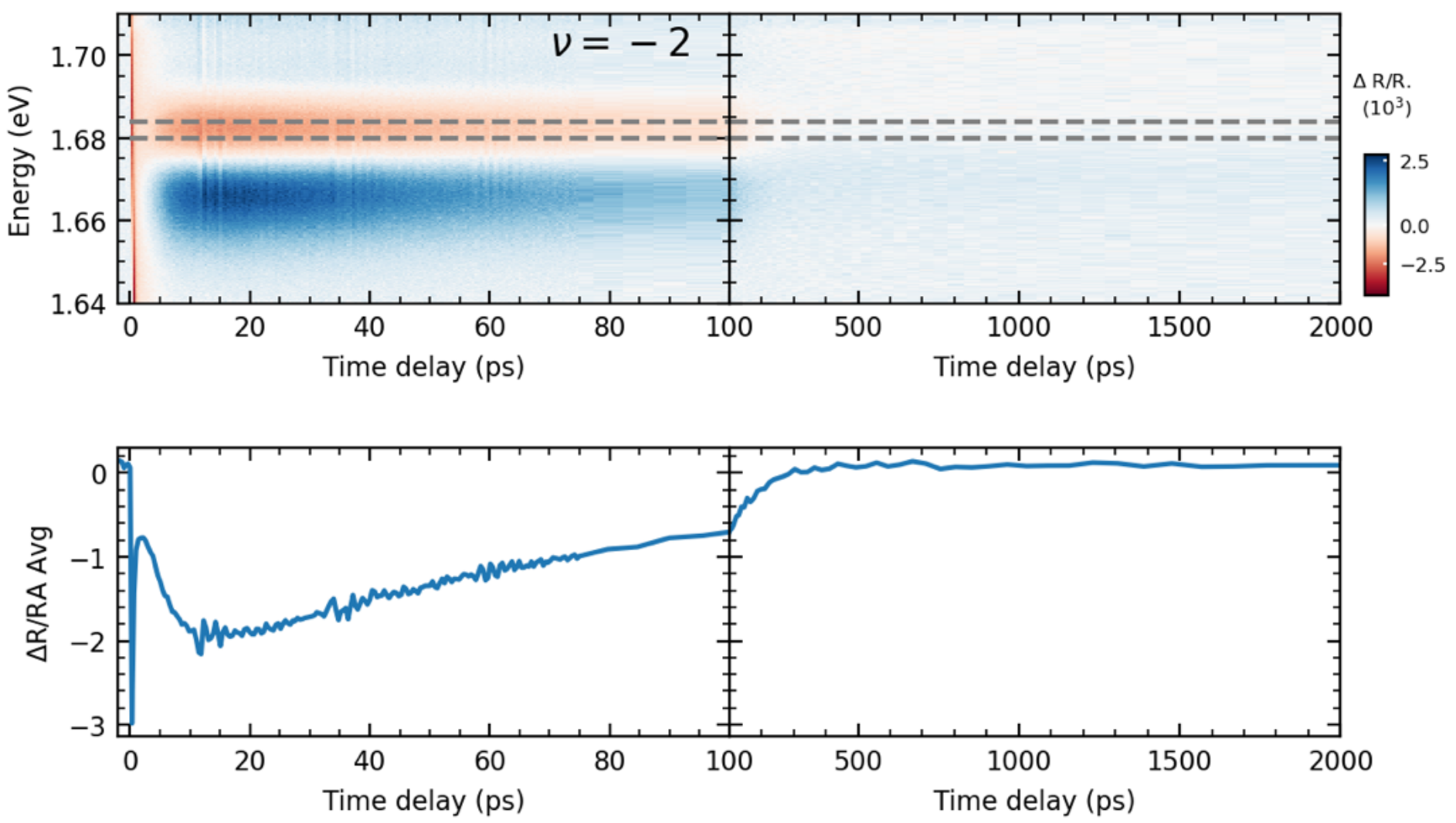

**Figure S6.** (Top) Transient reflectance spectra at a fixed carrier filling ($\nu_m$ = -2) as a function of pump-probe delays measured while probing the $WSe_2$ exciton. (Bottom) We show the cut at the minimum of the spectrum (integrated over a 10 meV range). This traces where measured with the same conditions as the transient gate scan shown in Figure 1 of the main text.

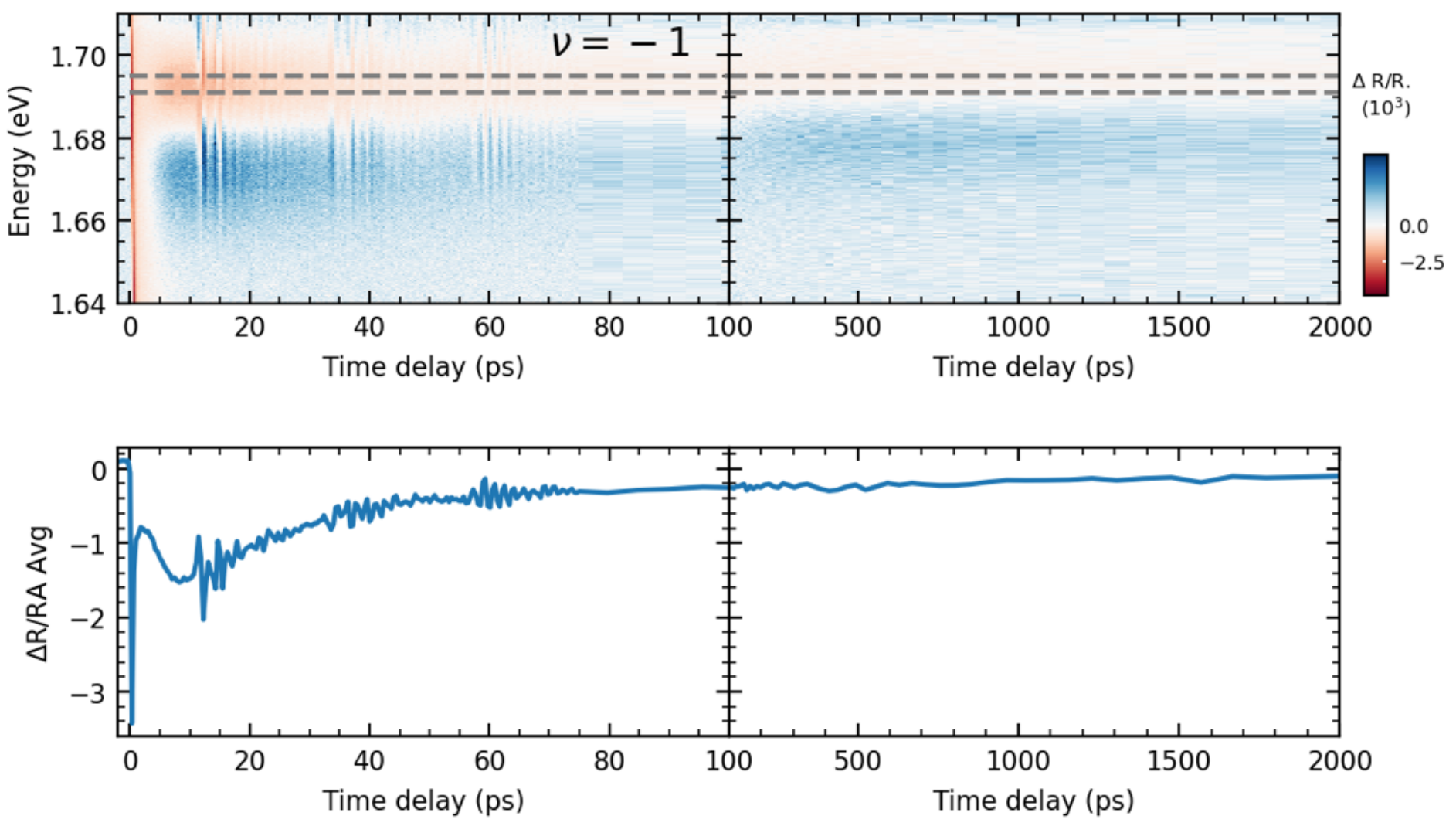

**Figure S7.** (Top) Transient reflectance spectra at a fixed carrier filling ($\nu_m$ = -1) as a function of pump-probe delays measured while probing the $WSe_2$ exciton. (Bottom) We show the cut at the minimum of the spectrum (integrated over a 10 meV range). This traces where measured with the same conditions as the transient gate scan shown in Figure 1 of the main text.

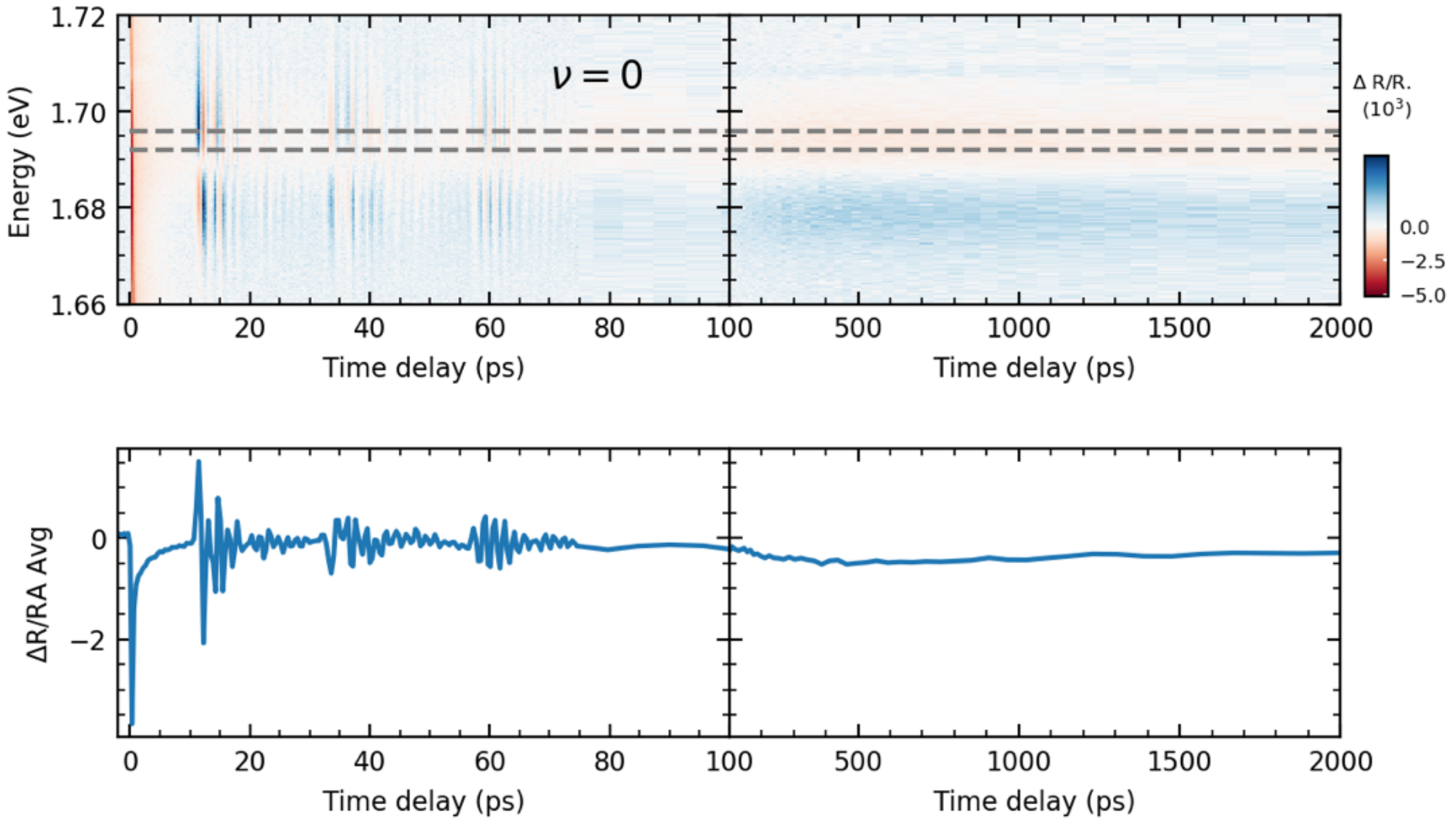

**Figure S8.** (Top) Transient reflectance spectra at a fixed carrier filling ( $\nu_m$ = 0) as a function of pump-probe delays measured while probing the $WSe_2$ exciton. (Bottom) We show the cut at the minimum of the spectrum (integrated over a 10 meV range). This traces where measured with the same conditions as the transient gate scan shown in Figure 1 of the main text.

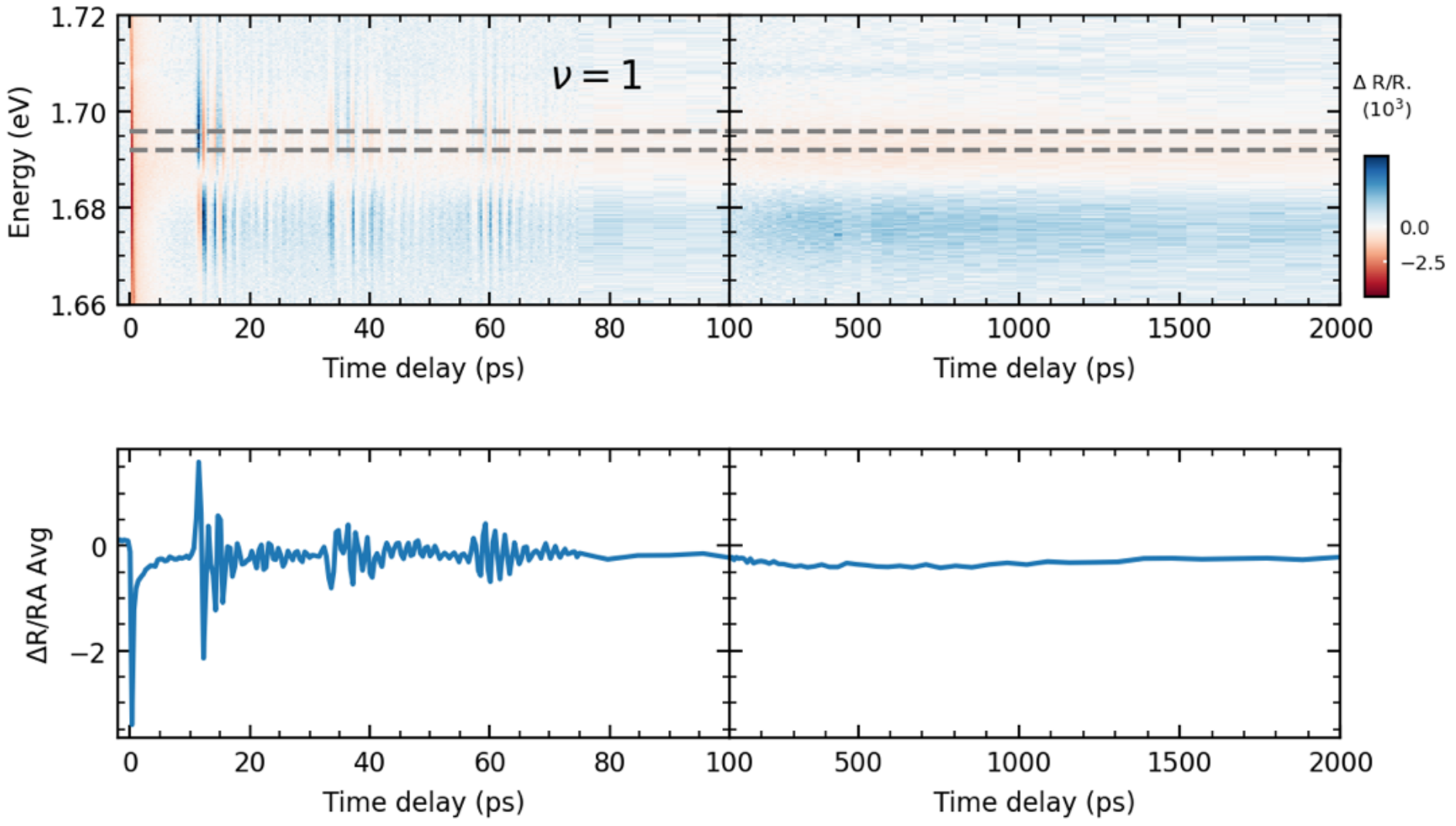

**Figure S9.** (Top) Transient reflectance spectra at a fixed carrier filling ( $\nu_m$ = 1) as a function of pump-probe delays measured while probing the $WSe_2$ exciton. (Bottom) We show the cut at the minimum of the spectrum (integrated over a 10 meV range). This traces where measured with the same conditions as the transient gate scan shown in Figure 1 of the main text.

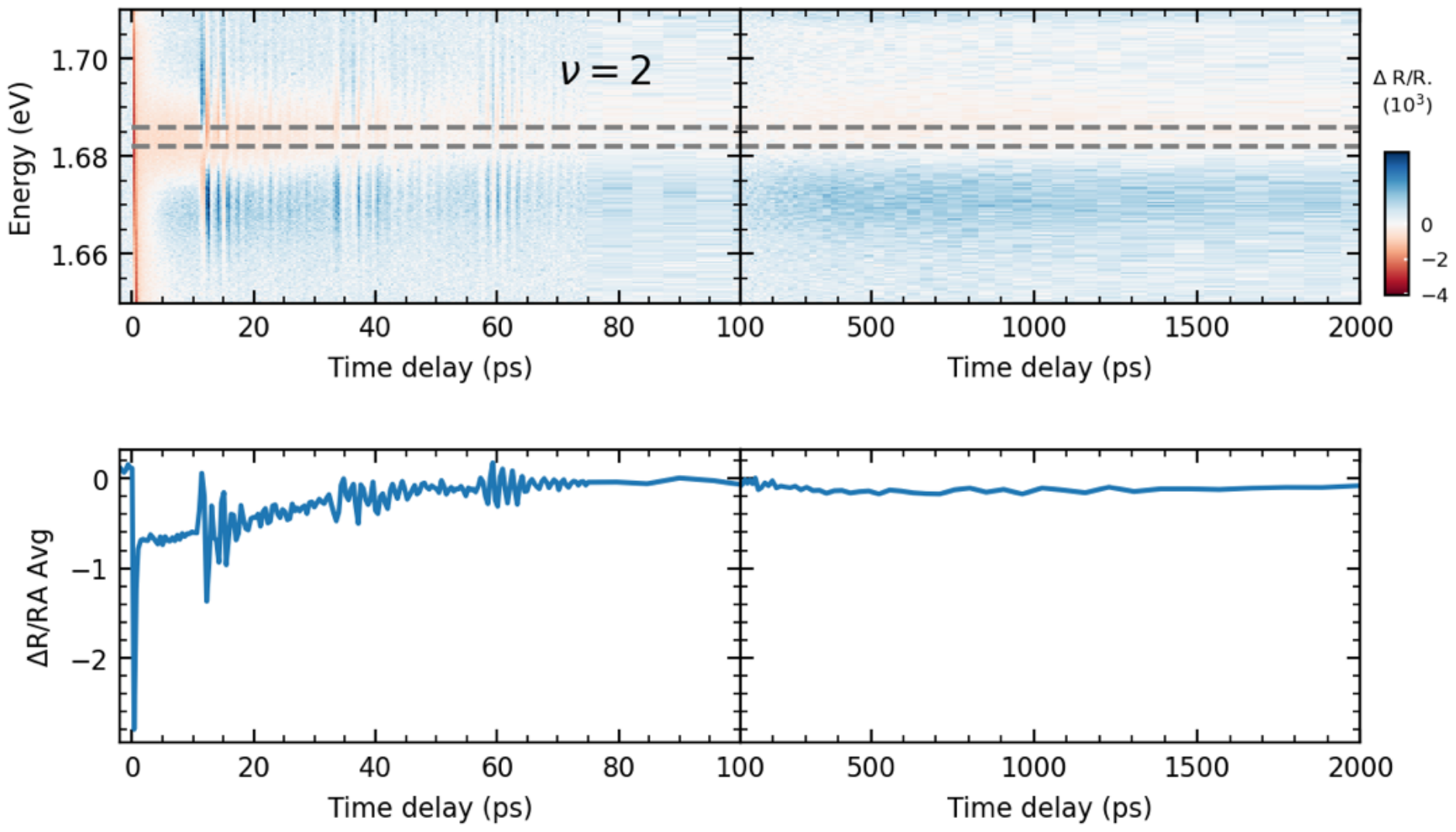

**Figure S10.** (Top) Transient reflectance spectra at a fixed carrier filling ($\nu_m = 2$) as a function of pump-probe delays measured while probing the $WSe_2$ exciton. (Bottom) We show the cut at the minimum of the spectrum (integrated over a 10 meV range). This traces where measured with the same conditions as the transient gate scan shown in Figure 1 of the main text.

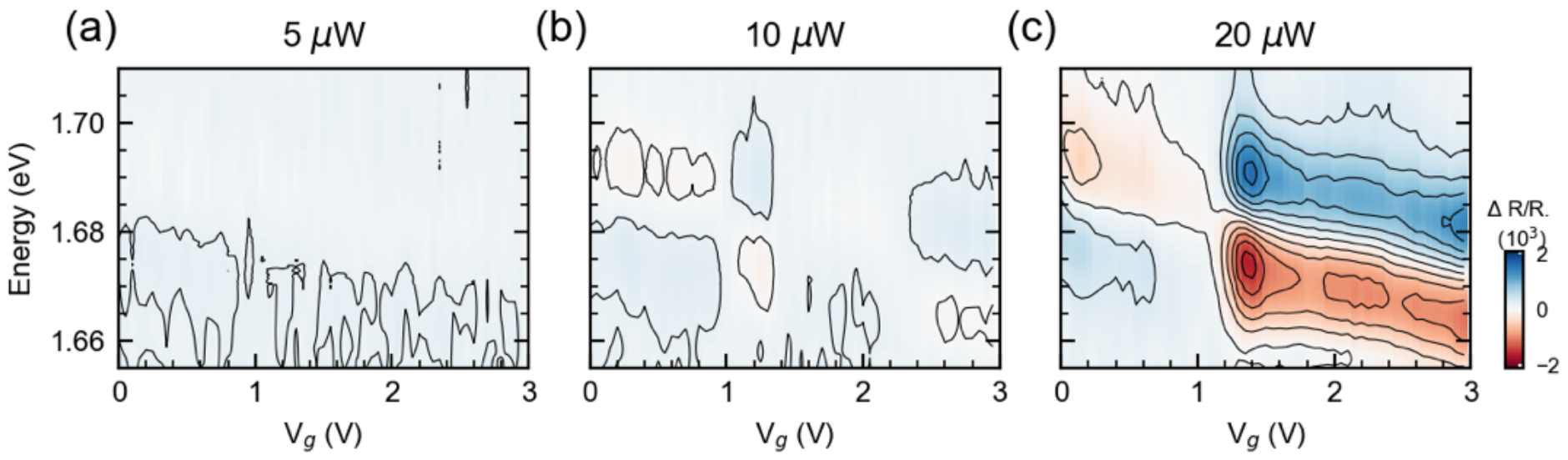

**Figure S11.** Fluence-dependent transient reflectance as a function of gate voltage at a pump-probe delay: Δt = -5 ps corresponding to panels (a) with a fluence of 106 μJ/μm$^2$, (b) with a fluence of 212 μJ/μm$^2$, and (c) with a fluence of 423 μJ/μm$^2$.

Dynamics for the $WSe_2/WS_2$ heterostructure at high-fluence 127 uJ/$cm^2$

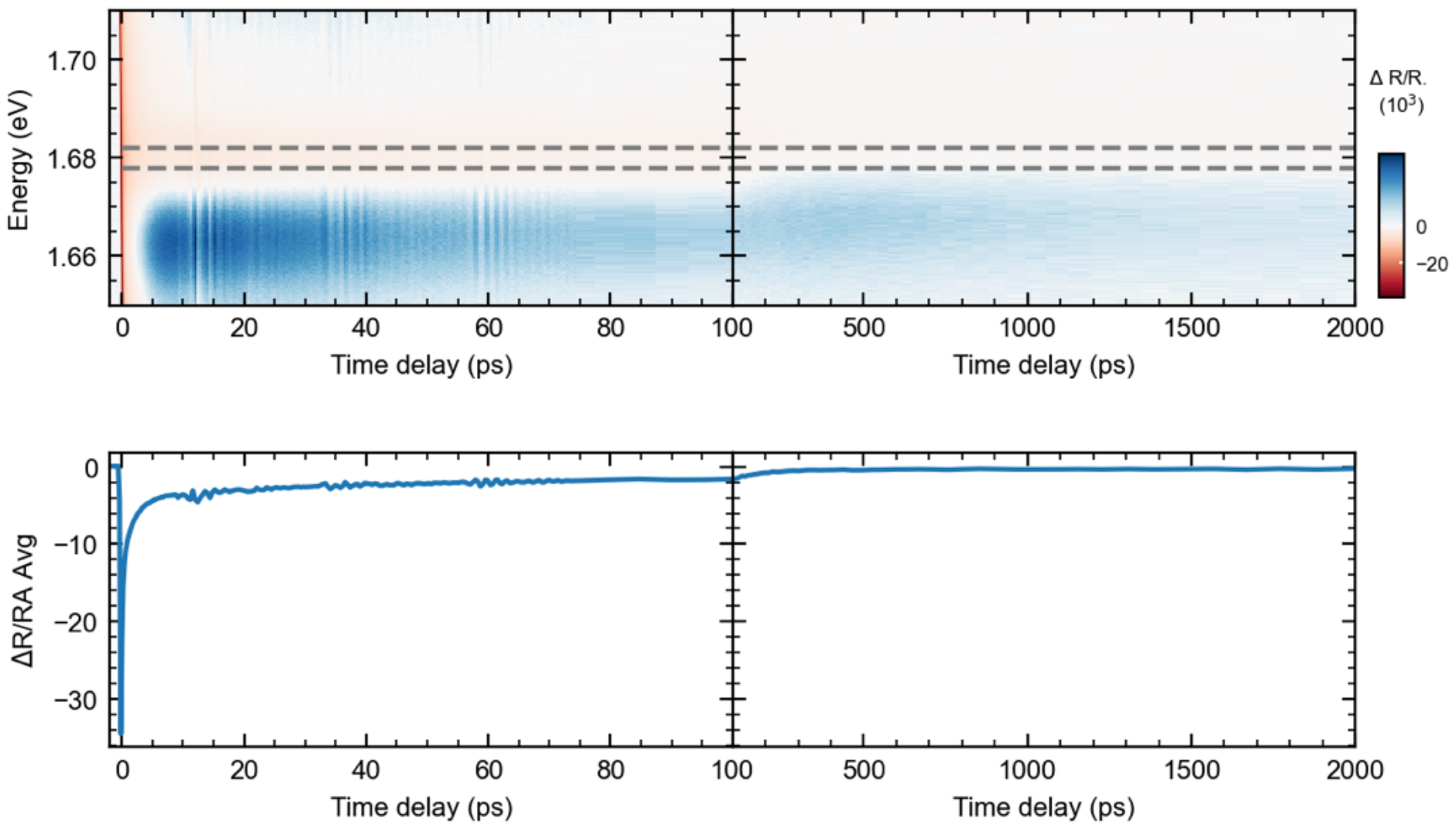

**Figure S12.** (Top) Transient reflectance spectra at a fixed carrier filling ( $\nu_m$ = -2) as a function of pump-probe delays measured while probing the $WSe_2$ exciton. (Bottom) We show the cut at the minimum of the spectrum (integrated over a 10 meV range). This traces where measured with the same conditions as the gate scan shown in Figure 2(e)-(h) of the main text.

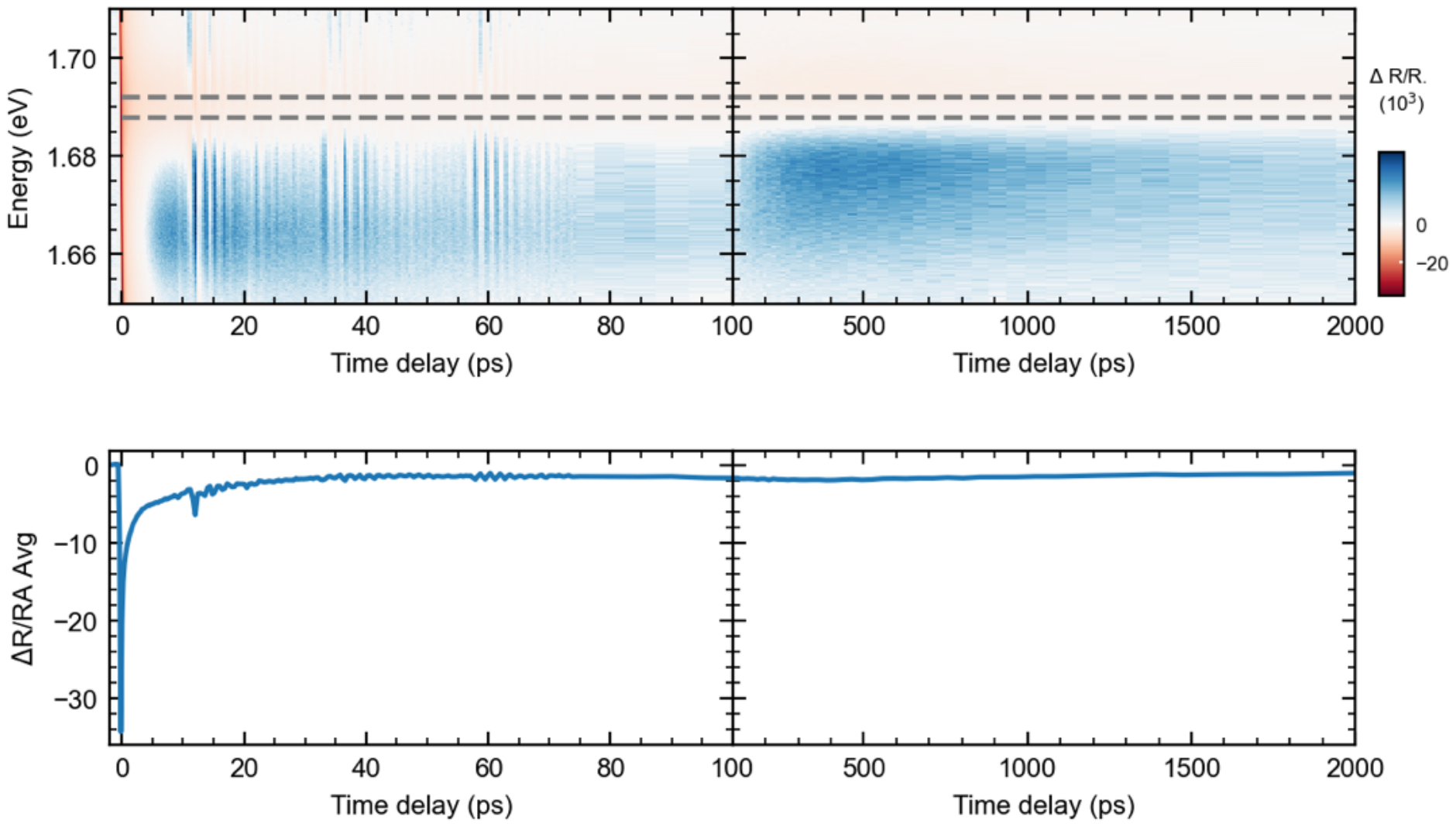

**Figure S13.** (Top) Transient reflectance spectra at a fixed carrier filling ( $\nu_m$ = -1) as a function of pump-probe delays measured while probing the $WSe_2$ exciton. (Bottom) We show the cut at the minimum of the spectrum (integrated over a 10 meV range). This traces where measured with the same conditions as the gate scan shown in Figure 2(e)-(h) of the main text.

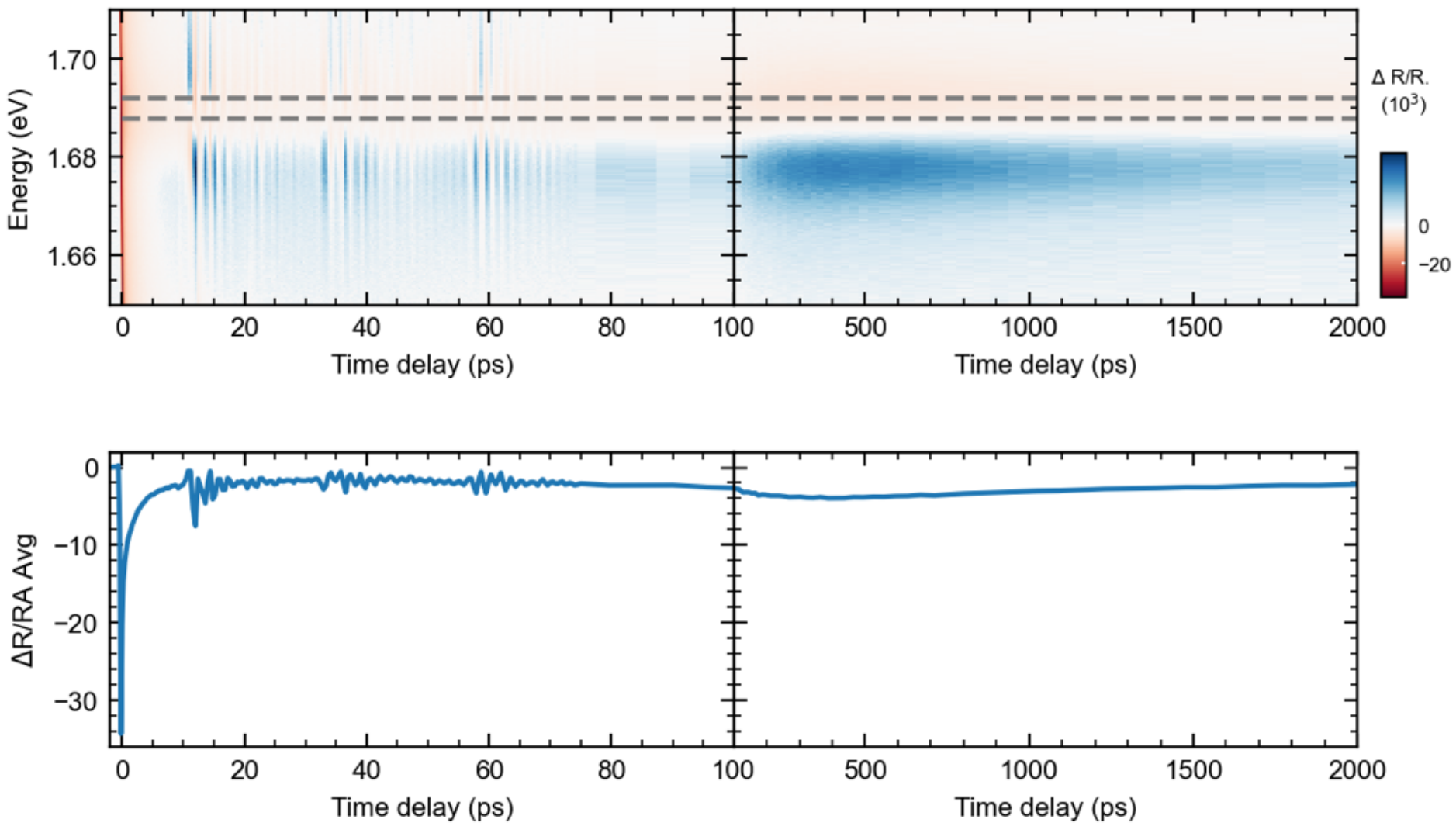

**Figure S14.** (Top) Transient reflectance spectra at a fixed carrier filling ( $\nu_m = 0$) as a function of pump-probe delays measured while probing the $WSe_2$ exciton. (Bottom) We show the cut at the minimum of the spectrum (integrated over a 10 meV range). This traces where measured with the same conditions as the gate scan shown in Figure 2(e)-(h) of the main text.

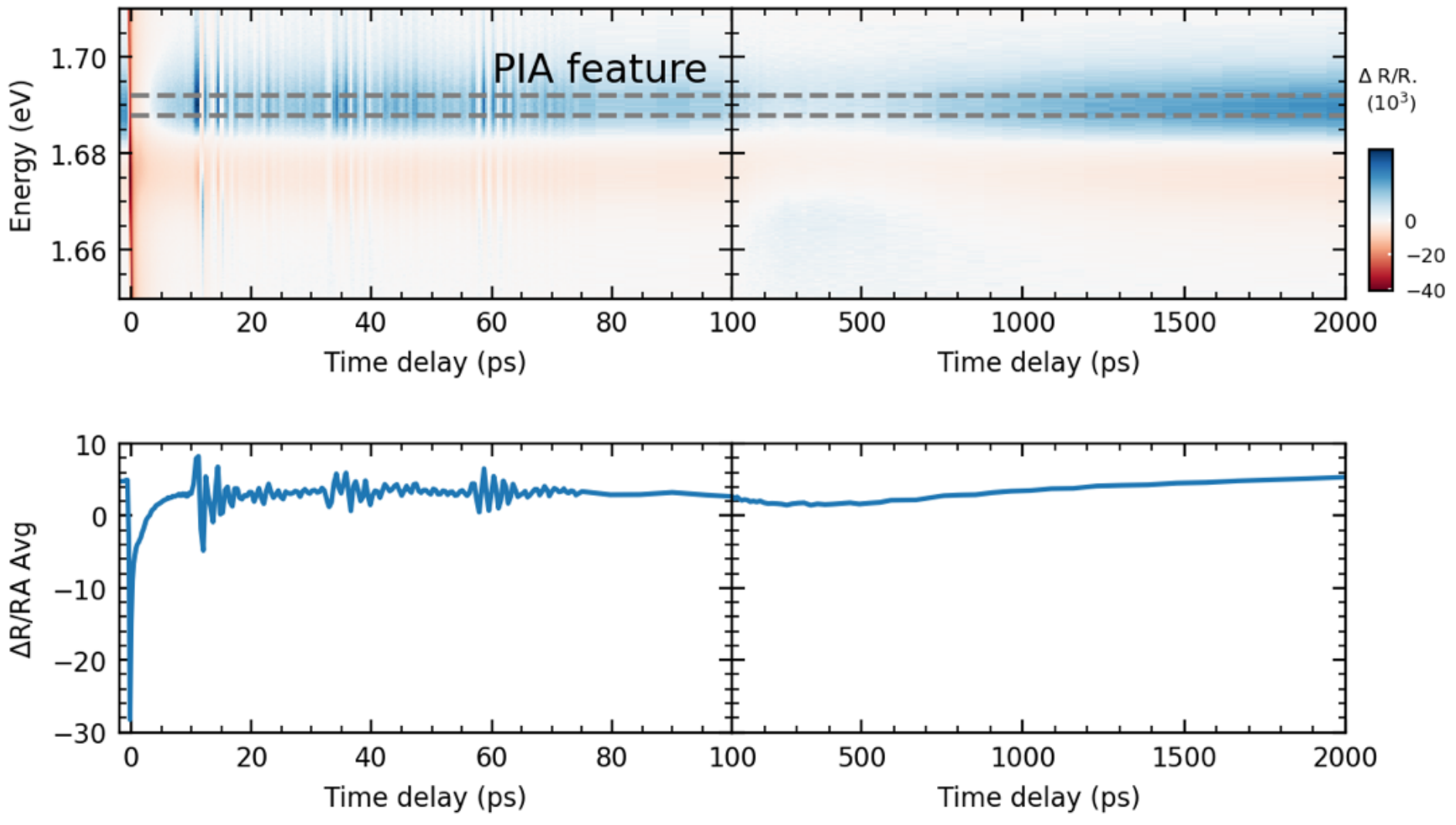

**Figure S15.** (Top) Transient reflectance spectra at a fixed carrier filling, consisting of the PIA-like feature, as a function of pump-probe delays measured while probing the $WSe_2$ exciton. (Bottom) We show the cut at the minimum of the spectrum (integrated over a 10 meV range). This traces where measured with the same conditions as the gate scan shown in Figure 2(e)-(h) of the main text.

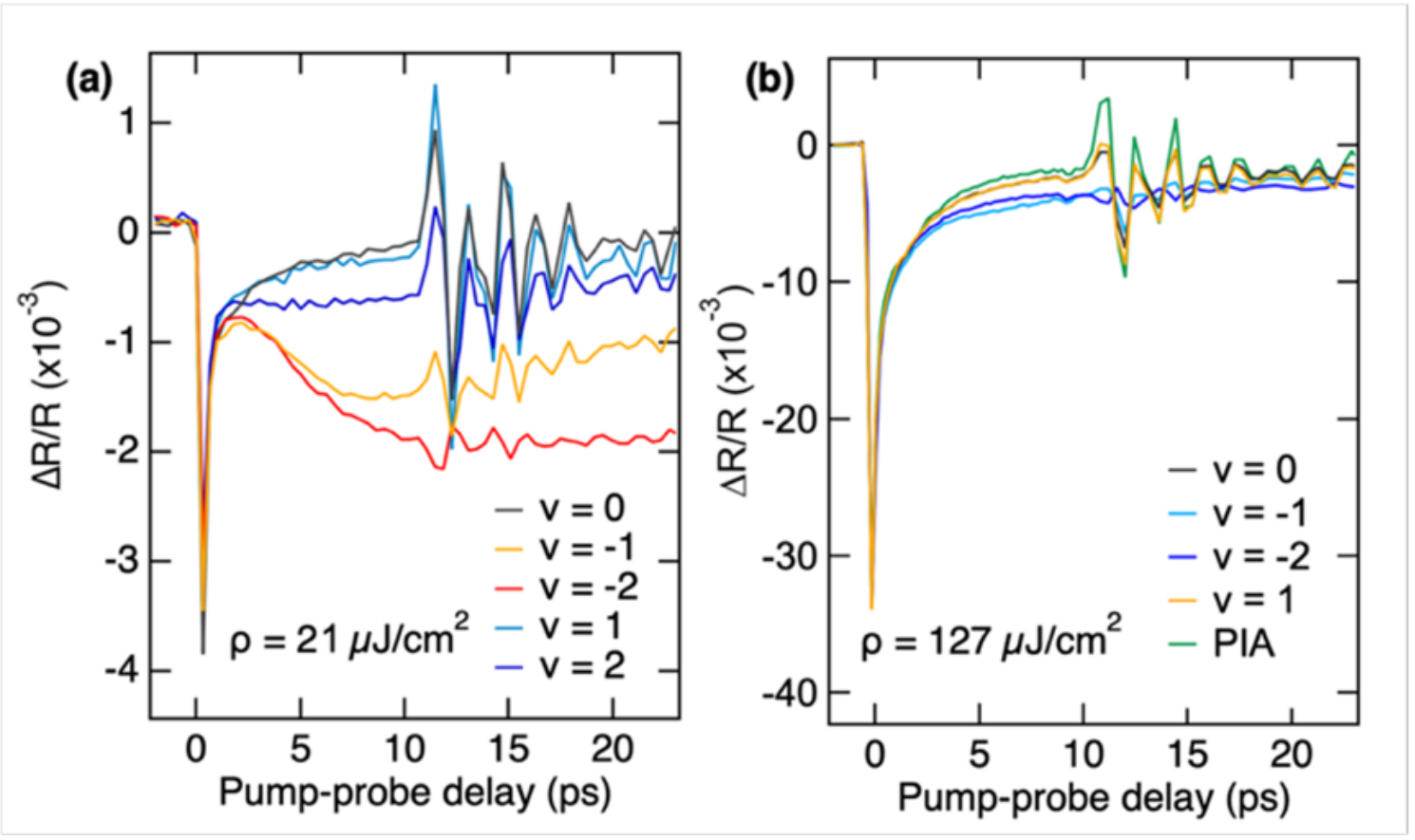

**Figure S15.** Transient reflectance in the short time window (≤ 24 ps) at the indicated doping levels for excitation pulse energy density (a) $\rho = 21$ μJ/cm$^2$, ν = 0, -1, -2, 1, 2; and (b) $\rho = 127$ μJ/cm$^2$, ν = 0, -1, -2, 1, and the PIA feature. All experiments carried out at a sample temperature of 10 K.

<u>Temperature-dependent transient reflection experiments in t$WSe_2$ device</u>

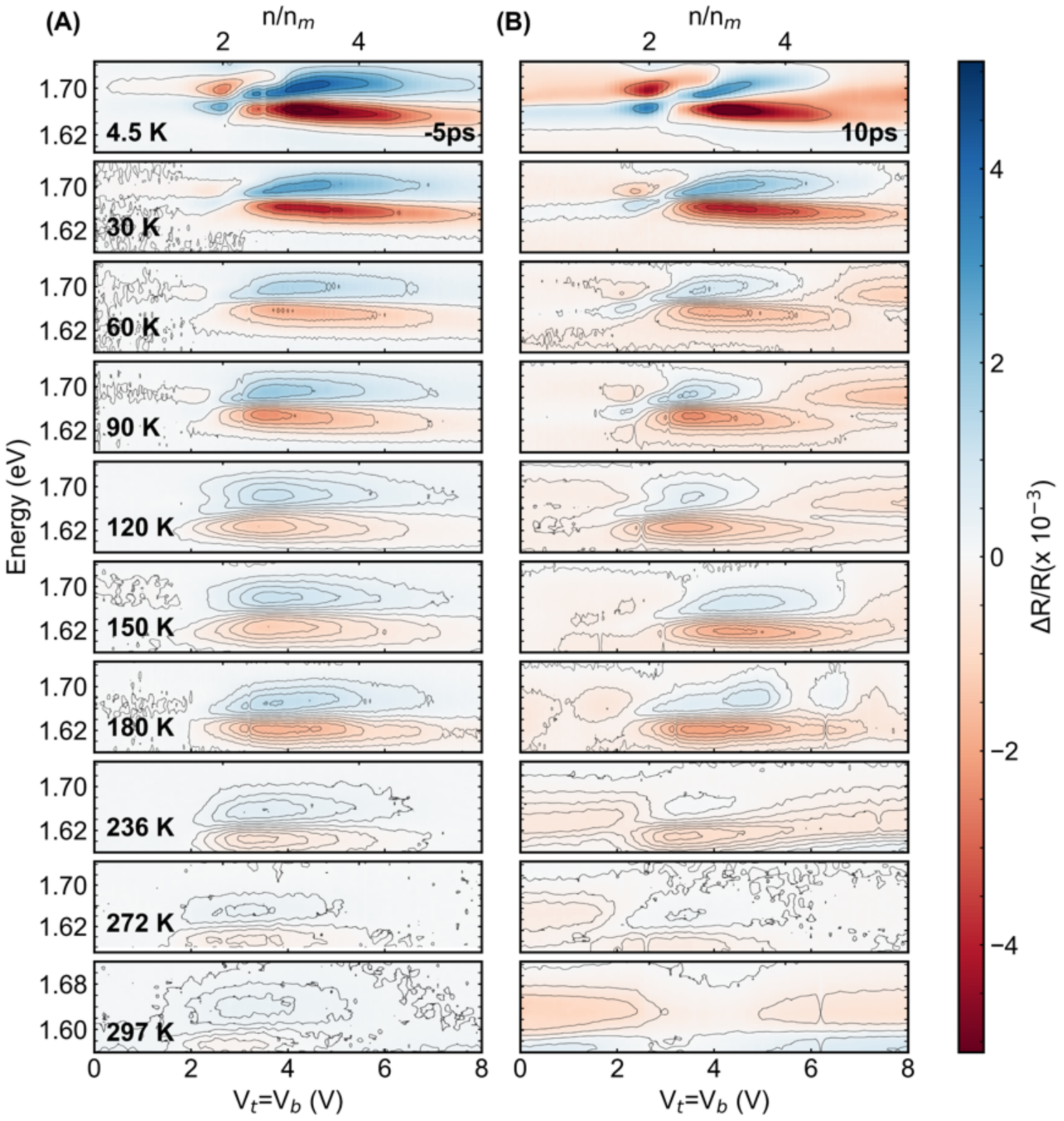

**Figure S16.** Temperature-dependent transient reflection response as a function of doping for two delay times between the pump and probe: 2.5 μs (-5 ps) (a) and 10 ps (b) at different temperatures from 4.5K to room temperature. The pump energy corresponded to 1.46 eV with a fluence of 127 μJ/μm$^2$, and the probe covered the first $WSe_2$ moiré exciton.